\documentclass{aa}
\usepackage{graphicx}
\usepackage{txfonts}
\usepackage{gensymb}
\usepackage{lscape}
\usepackage{xcolor}
\usepackage{caption,subcaption}
\usepackage[colorlinks = true,
            linkcolor = blue,
            urlcolor  = blue,
            citecolor = blue,
            anchorcolor = blue]{hyperref}

\begin{document}

  \title{Photometric determination of the mass accretion rates of pre-main-sequence stars\\  VIII. Recent star formation in NGC 299\thanks{Based on observations with the NASA/ESA \textit{Hubble} Space Telescope, obtained at the Space Telescope Science Institute, which is operated by AURA, Inc., under NASA contract NAS5-26555.}
  \thanks{Tables \ref{tab:catalogue_all} and \ref{tab:catalogue_PMS} are available in electronic form at the CDS via anonymous ftp to cdsarc.cds.unistra.fr (130.79.128.5) or via \url{https://cdsarc.cds.unistra.fr/cgi-bin/qcat?J/A+A/}}}

   \author{M. Vlasblom
          \inst{1}
          \and
          G. De Marchi\inst{2}
          }

   \institute{Leiden Observatory, Leiden University, 2300 RA Leiden, Netherlands\\
              \email{vlasblom@strw.leidenuniv.nl}
         \and
             European Space Research and Technology Centre, Keplerlaan 1, 2200 AG Noordwijk, Netherlands\\
             \email{gdemarchi@esa.int}
             }

   \date{Received October 19, 2022; accepted April 11, 2023}
\titlerunning{Recent star formation in NGC 299}
 
  \abstract
   {
   We studied the properties of the young stellar populations in the NGC 299 cluster in the Small Magellanic Cloud using observations obtained with the \textit{Hubble} Space Telescope in the $V, I$, and $H\alpha$ bands. We identified 250 stars with H$\alpha$ excess exceeding 5\,$\sigma$ and an  equivalent width of the H$\alpha$ emission line of at least 20\,\AA, which indicates that these stars are still undergoing active accretion and therefore represent bona fide pre-main-sequence (PMS) objects. For 240 of them, we derived physical stellar parameters such as the mass, age, and mass accretion rate by comparing the observed photometry with theoretical models. We find evidence that suggests the existence of two populations of PMS stars, one with a median age of around 25 Myr and the other about 50 Myr old. These ages are consistent with previously determined ages for the main population of the cluster. The average mass accretion rate for these PMS stars is $\sim 5 \times 10^{-9}$ M$_\odot$ yr$^{-1}$, which is comparable to the values found with the same method in  other low-metallicity, low-density clusters in the Magellanic Clouds, but is about a factor of three lower than those measured for stars of similar mass and age in denser Magellanic Cloud stellar regions. Our findings support the hypothesis that both the metallicity and density of the forming environment can affect the mass accretion rate and thus the star formation process in a region.\\
   \newline
   A study of the spatial distribution of both massive stars and (low-mass) PMS objects reveals that the former are clustered near the nominal centre of NGC 299, whereas the PMS stars are rather uniformly distributed over the field. Although it is possible that the PMS stars formed in situ in a more diffuse manner than massive stars, it is also plausible that the PMS stars formed initially in a more compact structure together with the massive stars and were later dispersed due to two-body relaxation. To explore this possibility, we studied the cluster's stellar density profile. We find a core radius $r_c\simeq 0.6$\,pc and a tidal radius $r_t\simeq 5.5$\,pc, with an implied concentration parameter $c \simeq 1$, suggesting that the cluster could be dispersing into the field.}


   \maketitle
%

\section{Introduction}\label{sec:intro}
In the current understanding of low-mass star formation (e.g. \citealt{hartmann2016}), the lifetime of a protostar is dominated by heavy accretion from the envelope surrounding the central star, and the infalling matter is thought to follow the star's magnetic field lines, a process known as magnetospheric accretion \citep{hartmann1998}. The accretion of matter is also thought to be episodic \citep{hartmann1996}, and the mass accretion rate can reach values as high as 10$^{-4}$ M$_\odot$ yr$^{-1}$ during an accretion outburst \citep{calvet2000}. Once the envelope is cleared out and the star reaches the pre-main-sequence (PMS) phase of its lifetime, the only accretion of matter onto the central star comes from the circumstellar disk. During this time, the mass accretion rate slows down significantly \citep{calvet2000}. Near-infrared studies of nearby small star-forming regions have shown accretion rates of about $10^{-8}$\,M$_\odot$\,yr$^{-1}$ at ages of $\sim 1$\,Myr and about an order of magnitude less around 10\,Myr. Correspondingly, the circumstellar disks of these PMS stars have been found to have a lifetime of around $2 - 3$\,Myr, with very few surviving after 10\,Myr \citep{haisch2001, hernandez2007}. \\
\newline
In this work, we are interested in the star formation history of the open star cluster NGC 299 in the Small Magellanic Cloud (SMC). To this end, we focus on characterising the PMS stars. Specifically, we focus on the masses, ages, and mass accretion rates of these objects to understand the timing and intensity of the star formation episodes. Usually, the mass accretion rate of young stellar objects is derived through spectroscopy, but this strongly limits what regions can be studied because of the high spatial resolution required in regions at a distance larger than about 1 kpc. As a result, most spectroscopic studies of young stellar objects focus on star-forming regions that are very nearby, which happen to be relatively small and have a metallicity comparable to the solar value. In contrast, star-forming regions in the Magellanic Clouds have a lower metallicity, of roughly $1/2$ to $1/8$\,Z$_\odot$ (see, e.g. \citealt{russell1992, geha1998, rolleston2002, lee2005, grocholski2009, choudhury2018}). To better understand the star formation process and the effects the environment can have on it, studying the young stellar objects outside the Milky Way and in the Magellanic Clouds in particular is very important, including in a cosmological context since the majority of stars in the Universe formed at z $\sim$ 2 \citep{madau2014}, when the prevailing metallicity was similar to that of the Magellanic Clouds.\\
\newline
A method for identifying PMS stars in extragalactic regions through photometry was proposed by \citet{demarchi2010}. Since excess H$\alpha$ emission is a characteristic signature of the presence of accretion onto a PMS star \citep{calvet2000}, \citet{demarchi2010} propose identifying PMS stars by looking at their $V - H\alpha$ colour excess. We follow this same method for our work, as explained in more detail in Sect. \ref{sec:PMSstars}. Interestingly, using this method allowed \citet{demarchi2010} (see also \citealt{demarchi2011, demarchi2013, demarchi2017}) to identify PMS stars that still show signs of accretion well beyond the typical 10 Myr expected lifetime of the circumstellar disks in nearby Galactic star-forming regions of low total mass, such as Taurus, $\rho$ Oph, Chamaeleon, and Cep OB2 (see, e.g. \citealt{sicilia-aguilar2010}).\\
\newline
It is presently not known how these disks can still feed their central stars for well over 20 Myr, but one possibility is the lower metallicity of the star-forming regions in the Magellanic Clouds. A lower metallicity of the gas in the circumstellar disk likely dampens the effect of the radiation pressure on the gas itself (see, e.g. \citealt{wilson2019}), which in turn could cause the circumstellar disk to be dispersed less efficiently, extending the duration of the accretion process. Indeed, \citet{demarchi2017} find that the mass accretion rate for objects of similar ages and masses is typically lower in lower-metallicity environments. However, \citet{demarchi2017} also note that the cluster NGC\,602 in the SMC seems to depart from the relation they derived. Even though its metallicity is similar to that of the other SMC cluster in their sample, NGC\,346, the accretion rate for stars of similar masses and ages is a factor of about two lower. They point out, however, that this cluster is located in a lower-density environment compared to the other clusters that they studied. Differences in the environments can be seen, for example, in the map of the dust surface density of the SMC \citep{utomo2019}. Following up on this, \citet{tsilia2023} studied the cluster NGC\,376, also located in a low-density region of the SMC, and found it to have mass accretion rates similar to those found in NGC\,602. These mass accretion rates were again systematically lower compared to NGC\,346, a SMC cluster in a higher-density environment, further suggesting that the environment of a cluster may have a significant impact on the mass accretion rate. Similar indications come from the works of \citet{biazzo2019} and \citet{carini2022}, who studied PMS objects in the Large Magellanic Cloud (LMC) clusters LH\,95 and LH\,91, respectively. They find that regions with lower dust densities also have lower mass accretion rates (and possibly lower gas densities). \\
\begin{figure}[h]
    \centering
    \includegraphics[scale=0.5]{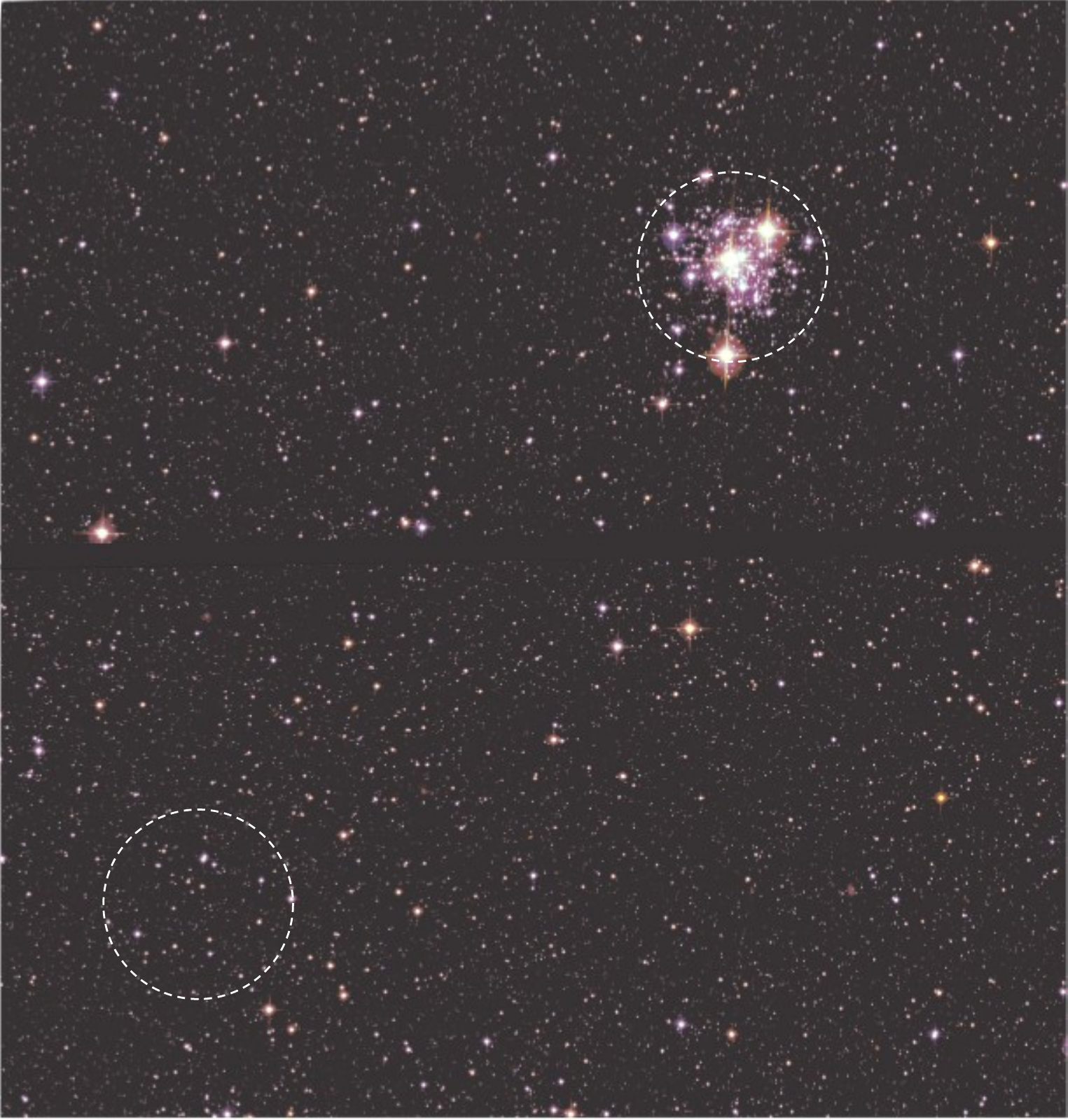}
    \caption{True colour image of NGC\,299. The red, green, and blue channels were assigned the I band, the median of the V and I bands, and the V band, respectively. The image spans roughly 50 pc per side. North is up, and east is to the left of north. The two circles drawn on the image indicate the cluster area (top right) and an offset region (bottom left).}
    \label{fig:Truecolour}
\end{figure}
\newline
As already mentioned, here we study the cluster NGC\,299, a young and relatively small cluster in the western outskirts of the SMC expected to have a relatively low density. The cluster age has been estimated in previous studies to be roughly 25 Myr \citep{piatti2008, glatt2010}. The age of the cluster, combined with the metallicity and density of the environment, makes it a suitable candidate for studying the impact of a cluster's environment on the process of accretion onto PMS stars.\\
\newline
The structure of the paper is as follows: in Sect. \ref{sec:photometry} we present the data and the photometric analysis. Section \ref{sec:CMD} is devoted to the colour-magnitude diagram (CMD) and an assessment of the extinction. In Sect. \ref{sec:PMSstars} we discuss how to identify PMS stars, and in Sect. \ref{sec:properties} we derive their physical parameters and associated uncertainties. Section \ref{sec:dynamical} is devoted to a study of the dynamical state of the cluster. A summary and conclusions are provided in Sect. \ref{sec:conclusions}.

\section{Observations and photometry}\label{sec:photometry}
For this work, images obtained with the \textit{Hubble} Space Telescope (HST) in three different photometric bands are used. The exposures taken in the F555W band (hereafter denoted as the `$V$' band) and in the F814W band (hereafter `$I$' band) were collected in 2004 September, as part of proposal number 10248 (principal investigator: A. Nota). These exposures were taken using the Wide Field Channel (WFC) of the Advanced Camera for Surveys (ACS). The exposure times for these images are 1858 and 1966\,s in the $V$ and $I$ bands, respectively. In addition to these long exposures, short exposures of 3\,s duration were also secured to allow for the correction of saturated stars. We also use exposures taken in the F656N band (hereafter denoted as the `$H\alpha$' band). These images were taken in 2013 August, as part of proposal number 13009 (principal investigator: G. De Marchi) and they were obtained from the UV/Visible (UVIS) channel of the Wide Field Camera 3 (WFC3). The combined $H\alpha$ image has an equivalent exposure time of 2394\,s. All downloaded images were already subjected to the standard HST calibration pipeline, which takes into account the bias offset, dark current, and flat-fielding correction, as well as the compensation for charge transfer efficiency and geometric distortion. All of this was done according to the ACS and WFC3 data handbooks (see \citealt{acshandbook2, wfc3handbook2}). A true-colour image of the region is shown in Fig. \ref{fig:Truecolour}.\\
\begin{figure}[h]
    \centering
    \includegraphics[scale=0.55]{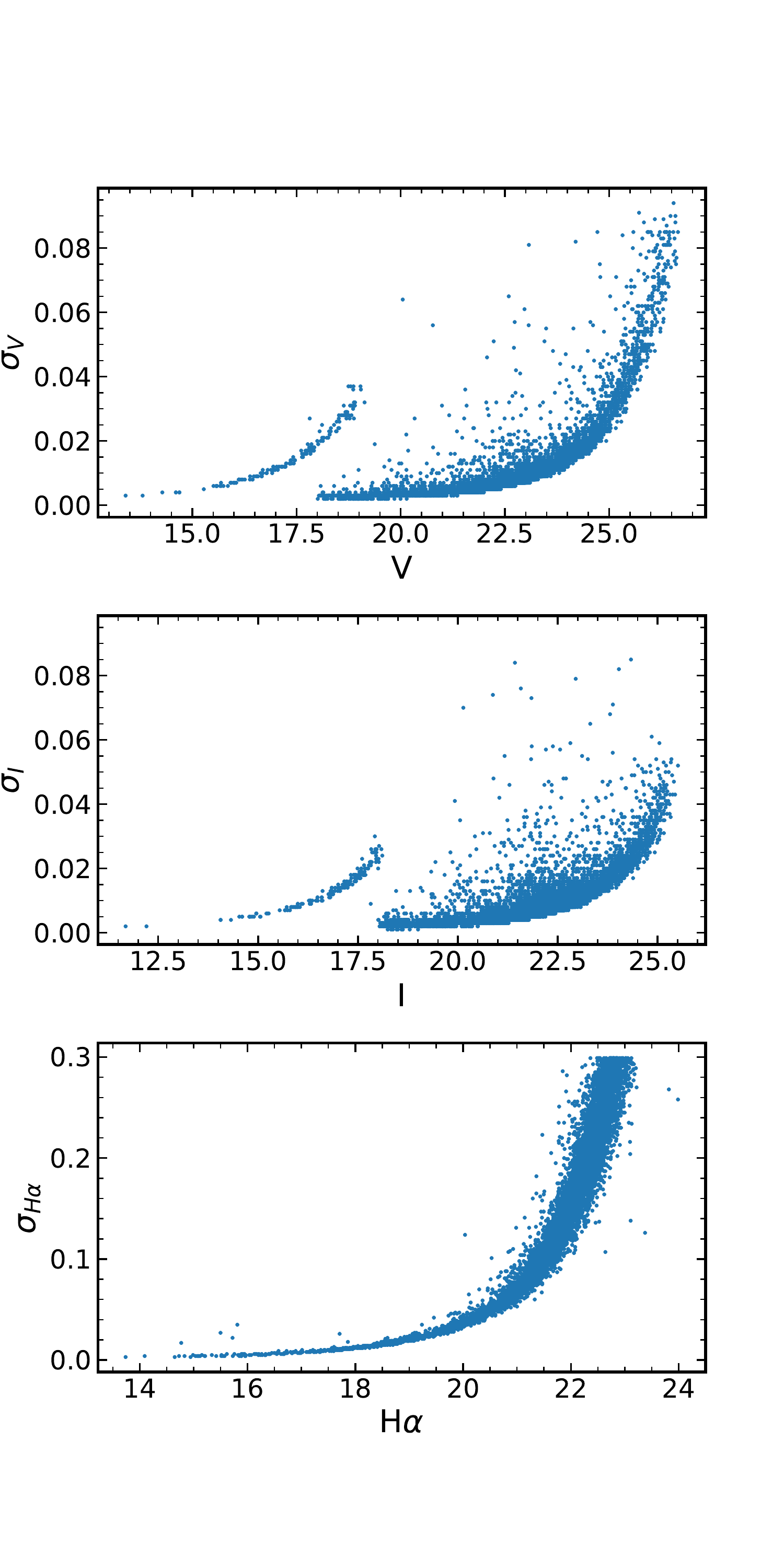}
    \caption{Magnitude uncertainties ($\sigma_V$, $\sigma_I$, $\sigma_{H\alpha}$) as a function of the magnitude for the three different bands. The $V$ and $I$ bands show two curves since the photometry for these stars was done in both short and long exposures. Note the difference in scale between the $V$ and $I$ bands with respect to the $H\alpha$ band.}
    \label{fig:Trumpet}
\end{figure}
\newline
To conduct the photometric analysis of the cluster, the DAOPHOT routine in IRAF (Image Reduction and Analysis Facility) was used \citep{tody1986}. We selected our objects by requiring them to be brighter than the local background by at least five times the local standard deviation of the background. Then, the magnitudes of these objects were determined using an aperture of 2 pixels in radius and a sky annulus that runs from 3 to 5 pixel. An aperture correction was applied to take into account the corresponding encircled energies and a zero-point correction was applied to properly calibrate the stellar magnitudes. For this, the ACS and WFC3 data handbooks were again followed. \\ 
\newline
Next, we narrowed down our dataset by selecting the stars with reliable photometry for the CMD we subsequently constructed (see Sect. \ref{sec:CMD}). We selected stars with a combined uncertainty in the $V$ and $I$ bands of less than $0.1$ mag, and once we added the $H\alpha$ photometry, we further narrowed this selection to contain only stars with an $H\alpha$ uncertainty less than $0.3$ mag. In total, 46617 objects are detected by our initial photometric analysis of the data, of which 7929 satisfy the conditions set on the uncertainties in the $V,$ $I$, and $H\alpha$ bands. Only these objects are considered in the rest of this work.\\  
\newline
We show in Fig. \ref{fig:Trumpet} the photometric uncertainty as a function of the magnitude in the various bands. The larger uncertainty in the $H\alpha$ band is due to the narrower filter, compared to the $V$ and $I$ bands. The two sets of points in the graphs for the $V$ and $I$ bands correspond to the two different exposure times. \\
\newline
The positions, magnitudes and uncertainties in the $V$, $H\alpha$, and $I$ bands for all 7929 stars in our final sample are contained in Table \ref{tab:catalogue_all}, of which only the first few lines are shown here. 

\begin{table*}[h]
    \centering
    \begin{tabular}{c c c c c c c c c}
        \hline
        \hline
         ID & RA  & Dec. & $I$ & $\delta\,I$ & $V$ & $\delta\,V$ & $H\alpha$ & $\delta\,H\alpha$  \\
         \hline
        1 & 0:53:24.0054 & -72:13:53.492 & 17.82 & 0.02 & 19.13 & 0.03 & 18.10 & 0.01 \\
        2 & 0:53:24.2016 & -72:13:52.578 & 21.98 & 0.01 & 22.65 & 0.01 & 21.68 & 0.14 \\
        3 & 0:53:24.0404 & -72:13:52.202 & 21.08 & 0.01 & 21.78 & 0.01 & 21.54 & 0.11 \\
        4 & 0:53:25.3437 & -72:13:52.041 & 21.89 & 0.01 & 22.51 & 0.01 & 22.25 & 0.13 \\
        5 & 0:53:23.9860 & -72:13:51.467 & 21.38 & 0.01 & 21.87 & 0.01 & 21.76 & 0.13 \\
        6 & 0:53:25.6099 & -72:13:51.432 & 22.49 & 0.01 & 23.34 & 0.01 & 22.37 & 0.20 \\
        7 & 0:53:25.1182 & -72:13:51.189 & 19.14 & 0.00 & 19.11 & 0.00 & 19.08 & 0.02 \\
        8 & 0:53:24.7255 & -72:13:51.185 & 20.31 & 0.01 & 21.37 & 0.01 & 20.63 & 0.06 \\
        9 & 0:53:24.6611 & -72:13:51.214 & 18.45 & 0.01 & 19.67 & 0.01 & 18.66 & 0.02 \\
        10 & 0:53:24.5414 & -72:13:50.998 & 23.37 & 0.02 & 24.12 & 0.02 & 22.69 & 0.24 \\
        \hline
        \hline
    \end{tabular}
    \caption{Position, magnitudes, and uncertainties of the 7929 selected sources. {The RA and Dec. positions are those in the V band frame.} Only the first few lines are shown; the complete catalogue is available at the CDS.}
    \label{tab:catalogue_all}
\end{table*}

\section{Colour-magnitude diagram}\label{sec:CMD}
\begin{figure}[h]
    \centering
    \includegraphics[scale=0.45]{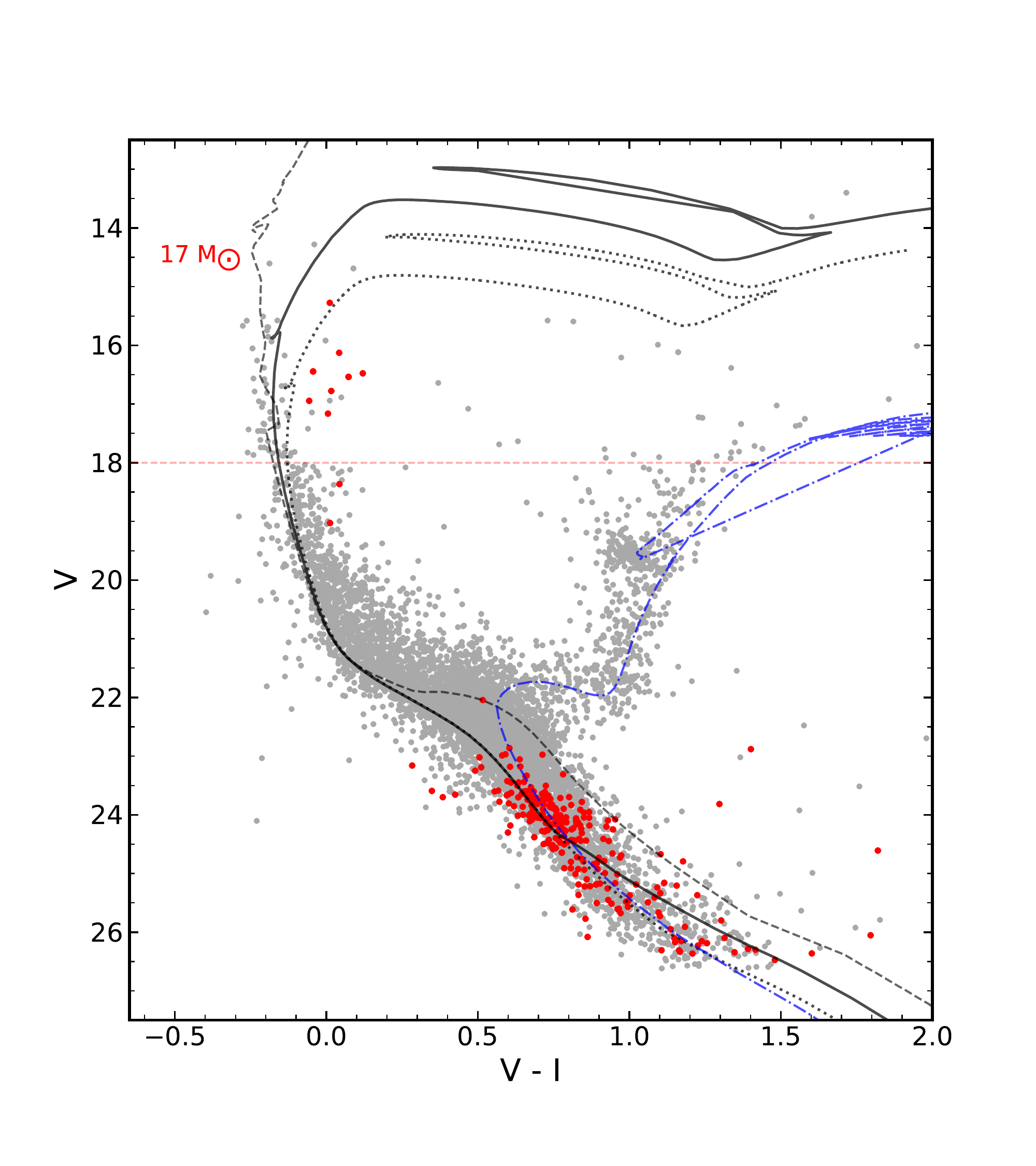}
    \caption{CMD of NGC\,299. Red dots indicate the stars with $H\alpha$ excess emission. Isochrones for ages 10 Myr, 30 Myr, 60 Myr, and 4 Gyr are also shown (dashed, solid, dotted, and dot-dashed lines, respectively). The approximate mass of the brightest young star still close to the MS is also indicated.}
    \label{fig:CMD}
\end{figure}

\begin{figure*}[ht]
    \centering
    \makebox[\textwidth][c]{\includegraphics[scale=0.5]{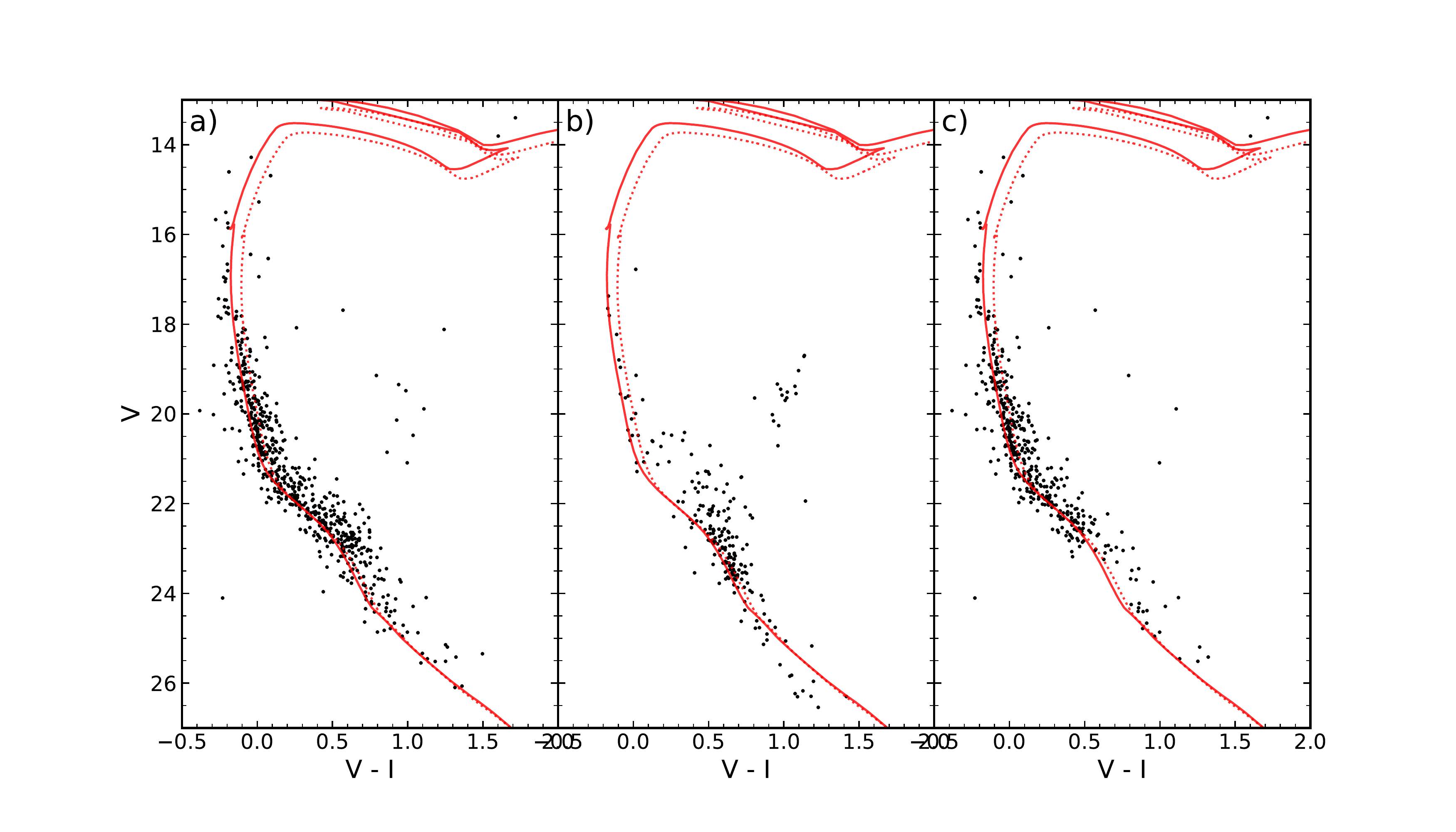}}
    \caption{Statistical subtraction of the cluster and field CMDs. Panel a) shows the CMD for stars projected within 15$\arcsec$ of the cluster centre. Panel b) shows the CMD for off-cluster stars inside a region of 15$\arcsec$ radius at the south-east corner of the region. Panel c) shows the statistical subtraction of the two CMDs. In all figures, the isochrone for 30 Myr is plotted (solid red line), as is the same isochrone translated to account for twice the extinction caused by the Milky Way galaxy (dotted red line).}
    \label{fig:Statsub}
\end{figure*}

The resulting CMD is shown in Fig. \ref{fig:CMD}, where all 7929 stars in our finally selected sample are marked in grey. Stars with $H\alpha$ excess emission above a certain threshold (PMS candidates) are shown in red. These stars, their detection method, and their physical properties are discussed in Sects. \ref{sec:PMSstars} and \ref{sec:properties}. Stars brighter than $V=18$ (see the thin horizontal line in Fig. \ref{fig:CMD}) were measured on the short exposures, to avoid the effects of saturation.\\
\newline
The CMD reveals a prominent main sequence (MS) that contains quite a few bright, massive stars, as well as many old stars in the red giant branch (RGB), along which the red clump (RC) can also be identified. The RC is slightly extended, revealing the presence of some differential extinction in this field (see, e.g. \citealt{demarchi2014, merica2017}), which we discuss later. \\
\newline
The isochrones shown in Fig. \ref{fig:CMD} correspond to ages of 10, 30, and 60 Myr (the dashed, solid, and black dotted lines, respectively), as well as 4 Gyr (the dot-dashed blue line) for the RGB. The isochrones, from the Padova group (see, e.g. \citealt{marigo2017}, and references therein) correspond to metallicity $Z=0.004$, as appropriate for the SMC (see, e.g. \citealt{russell1992, rolleston2002, lee2005}), and already include the reddening contribution of the Milky Way along the line of sight ($A_V = 0.18$; \citealt{carlson2017, schmalzl2008}). The 30 Myr isochrone appears to agree well with the approximate age of the cluster. Concerning the PMS stars, we can already see that there is some age spread in the CMD, even after taking into account the photometric uncertainty, with PMS candidates both younger and older than 30 Myr. We examine this in more detail in Sect. \ref{subsec:mass_age}. The massive stars also reveal a spread in age. The bluest objects are compatible with ages younger than 30 Myr and most likely closer to 10 Myr. In that case, the most massive of these stars, at $V\simeq 14.5$ would have a mass of $\sim 17$\,M$_\odot$. On the other hand, red supergiants appear to have an age close to $\sim 60$\,Myr. \\
\newline
To assess the effects of differential extinction, one would ideally remove the contribution of the field stars in the image from the CMD. However, due to the fact that no dedicated offset region has been observed for this purpose, completely removing this contamination is difficult. It is completely possible that different populations of stars are interspersed throughout the image. This is also discussed further in Sect. \ref{subsec:spatial}, where we show, for example, that the massive stars are distributed throughout the image (although most of them are still near the cluster). This is not surprising, as we just concluded from the CMD that their approximate age is 30 Myr, so they can indeed have moved away from the cluster itself. Thus, we performed a statistical subtraction where we subtract a sample of off-cluster stars from the CMD. As our cluster sample, we selected all stars within a radius of 15$\arcsec$ around the cluster centre, and all stars within a region of the same size but at the opposite corner of the image as our off-cluster sample. The location and approximate size of these regions is indicated in Fig. \ref{fig:Truecolour}. \\
\newline
We constructed a CMD for both the cluster and off-cluster samples (shown in panels a) and b) of Fig. \ref{fig:Statsub}) and then performed a statistical subtraction. We divided the CMDs into smaller cells and counted the number of stars in each cell in the cluster CMD and off-cluster CMD. The cells, which are identical in both CMDs, are defined in such a way that there are enough objects in each of them for the cluster CMD, in order to limit the effects of small number statistics. The statistically-subtracted CMD is shown in panel c) of Fig. \ref{fig:Statsub}. Along with the CMDs, two isochrones are shown, one for an age of 30 Myr (same as in Fig. \ref{fig:CMD}), and the other for the same age but shifted by an amount corresponding to twice the extinction produced by the Milky Way galaxy in the direction of the SMC, namely A$_{\text{V}}=0.18$, E(V-I) = 0.04 (e.g. \citealt{sabbi2011}). \\
\newline
As panel c) of Fig. \ref{fig:Statsub} shows, the cluster MS is rather narrow. Its spread is contained mostly within the separation of the two isochrones, indicating that the extinction experienced by the cluster is likely no more than twice the foreground extinction caused by the Milky Way. This corresponds roughly to an uncertainty of no more than $0.04$ in colour and $0.12$ in the $V$ magnitude, which in this analysis can be considered as negligible for our purposes. So, differential extinction is not considered further in this work.

\section{Identifying PMS stars}\label{sec:PMSstars}
\begin{figure*}[ht]
    \centering
    \makebox[\textwidth][c]{\includegraphics[scale=0.6]{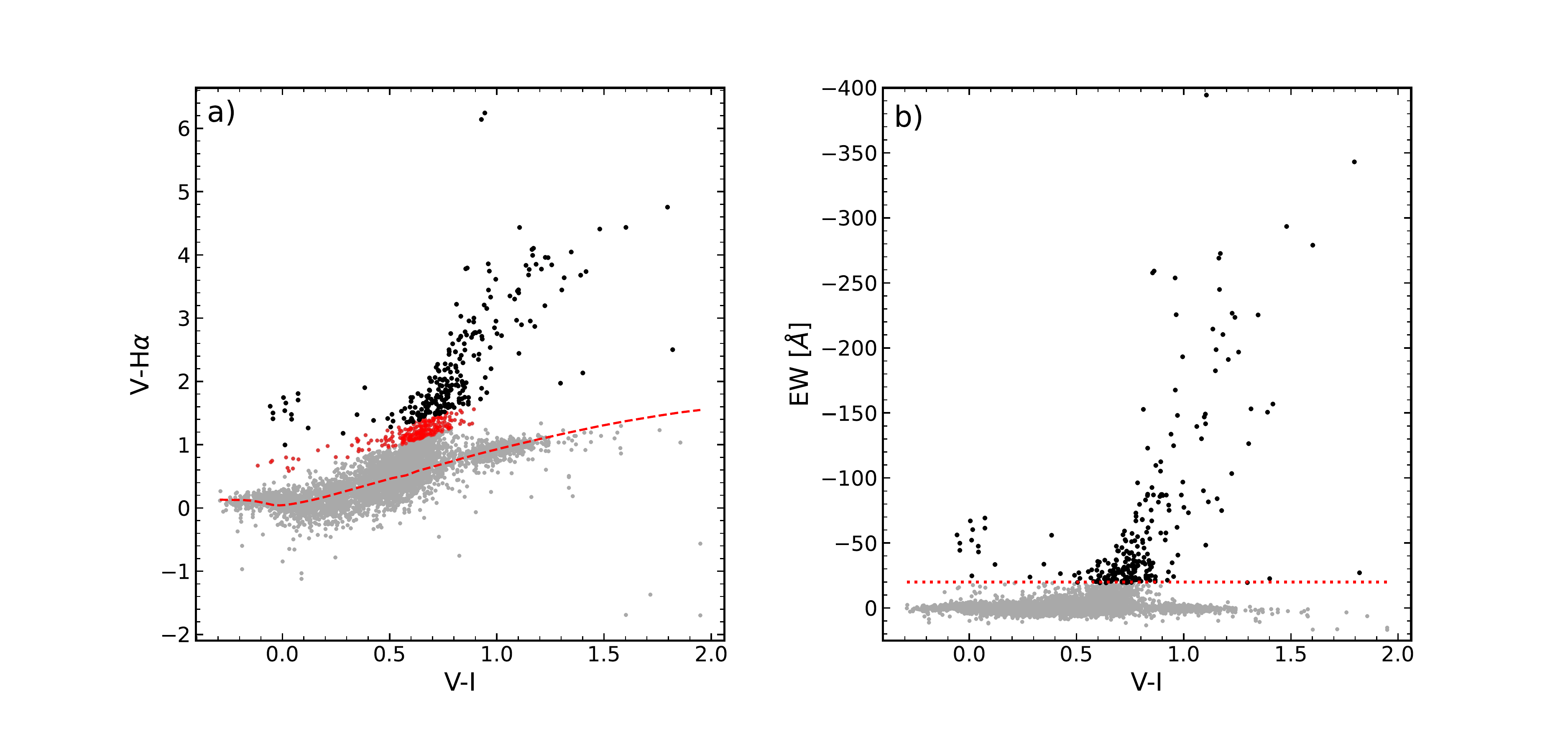}}
    \caption{Diagrams showing the selection of the PMS stars. Panel a): CCD showing the $V - I$ and $V - H\alpha$ colours of the stars. The dashed red line indicates the theoretical colour, following the \citet{bessell1998} models. Black dots indicate the stars with $H\alpha$ excess larger than 5\,$\sigma$ and an EW smaller than -20\,\AA. Red dots indicate stars with $H\alpha$ excess larger than 5\,$\sigma$ but which have an EW larger than -20\,\AA. Following our EW selection criterion, these stars are thus not included in our sample of bona fide PMS stars. Panel b): EW in $H\alpha$ plotted against the $V - I$ colour of the stars. The black dots indicate the stars with $H\alpha$ excess larger than 5\,$\sigma$ and an EW smaller than -20\,\AA,  and the red dotted line indicates an EW of $-20$\,\AA. }
    \label{fig:CCD}
\end{figure*}

The selection of PMS star candidates is based on the identification of objects with $H\alpha$ excess emission. This is illustrated in the colour-colour diagram (CCD) shown in panel a) of Fig. \ref{fig:CCD}, where the $V-H\alpha$ colour is plotted against the $V-I$ colour. The red dashed line in this figure shows, as a reference, the expected colours of normal MS stars, as per the models of \citet{bessell1998} for metallicity $M/H=-0.5$. In the rest of the paper we refer to this as the `reference curve'. Stars with $H\alpha$ excess emission stand out in a graph of this type, since they have $V-H\alpha$ colours larger than those of the reference curve at the same $V-I$ colour (see \citealt{demarchi2010, demarchi2011}, for more details) We initially selected the stars with an $H\alpha$ excess larger than 5\,$\sigma$ to be our bona fide PMS stars, where $\sigma$ is the combined uncertainty in the $V-H\alpha$ colour. \\
\newline
It should be noted that the measurements in the $V$ and $I$ bands are contemporaneous with each other, but not with the $H\alpha$ measurements. However, the $H\alpha$ excess emission observed is far too large to be caused by photometric variability of the continuum. Variations of this type are rare, so it is unlikely that so many of such variable stars are all present in our sample. In fact, when searching our field of view in the Optical Gravitational Lensing Experiment (OGLE) survey\footnote{\url{ogledb.astrouw.edu.pl/~ogle}}, only five variable stars can be found. Two of these are Cepheid variables with amplitudes of 0.5 and 0.3 magnitudes, respectively, and the other three are long-period variables with amplitudes between 0.016 and 0.006 mag. So, none of the known variable stars in the field of NGC 299 have the required variability to show an $H\alpha$ excess as large as what we see in our data.\\
\newline
To improve the selection criterion, we also derived the equivalent width (EW) of the $H\alpha$ emission line that corresponds to the observed colour difference between the $V-H\alpha$ value of a star and that of the reference curve for the same value of $V-I$ (which in practice provides the level of the continuum). In the case of narrow-band photometry, following \citet{demarchi2010} and using the rectangular width (RW) of the filter  (which is conceptually similar to the EW and fully defined by the properties of the filter), the EW of the line probed by the filter can be written as\begin{align}
    \label{eq:EW}
    \text{EW} = \text{RW} \cdot \left(1 - 10^{0.4\Delta\text{H}\alpha} \right),
\end{align}
where
\begin{align}
    \label{eq:Delta}
    \Delta\text{H}\alpha = \left(\text{V} - \text{H}\alpha\right)_{\text{obs}} - \left(\text{V} - \text{H}\alpha\right)_{\text{ref}}.
\end{align}
Here, $\Delta\text{H}\alpha$ is the $H\alpha$ excess, the difference between the observations (subscript `obs') and the reference curve (subscript `ref'). The RW is specific to the F656N filter mounted on the WFC3 instrument and corresponds to 17.679. The resulting EW as a function of the $V-I$ colour can be seen in panel b) of Fig. \ref{fig:CCD}. Since the lines are in emission, we assigned a negative value to them, following the standard convention. Placing a condition on the EW, in addition to the one on the strength of the $H\alpha$ excess emission, allows us to avoid some slowly rotating stars and brown dwarfs with active chromospheres. It was shown by \citet{white2003} that such stars typically have an absolute EW of less than 10{\AA}, while classical T Tauri stars have a much larger EW. Since we do not have spectra for our candidate PMS stars, our determination of the EW is necessarily less accurate. Therefore, we adopt a more conservative threshold than that suggested by \citet{white2003}, namely EW$<-20$\,\AA. This also allows us to exclude stars with active chromospheres from our sample. When combining this criterion together with our 5\,$\sigma$ selection on the $H\alpha$ excess, this selection yielded 250 objects, which we consider bona fide PMS candidates.\\
\newline
Figure \ref{fig:CCD} also reveals that most stars with $H\alpha$ excess are found in a sort of `plume' starting at $V-I \ga 0.5$, but there is a small group of 10 much bluer stars. In the CMD in Fig. \ref{fig:CMD}, these objects are located at magnitude $V<20$, next to the upper MS, whereas the other stars are much fainter. This small group of bright stars could be Be type stars, characterised by $H\alpha$ emission due to their intense winds (see, e.g. \citealt{porter2003}). We did not consider these objects further in this work, where we concentrate on the remaining 240 stars with $H\alpha$ excess.

\section{Physical properties of PMS stars}\label{sec:properties}
In the following sections, we address the properties of the PMS stars that were identified in Sect. \ref{sec:PMSstars}. This identification is based on a detection of $H\alpha$ excess emission at the 5\,$\sigma$ level and an EW $< -20$\,\AA\, at the time that the observations were taken. Thus, it is possible and, in fact, likely that the 240 objects identified are but a fraction of the overall PMS population in this field, owing to the large variations in the PMS stars' $H\alpha$ emission over hours and days (e.g. \citealt{Fernandez1995, Smith1999, Alencar2001}). Nevertheless, these are the PMS objects that we can securely identify in this field and we expect them to represent a statistically significant sample of the overall population of PMS stars. 
\subsection{Masses and ages}
\label{subsec:mass_age}
\begin{figure*}[ht]
        \centering
        \makebox[\textwidth][c]{\includegraphics[scale=1.3]{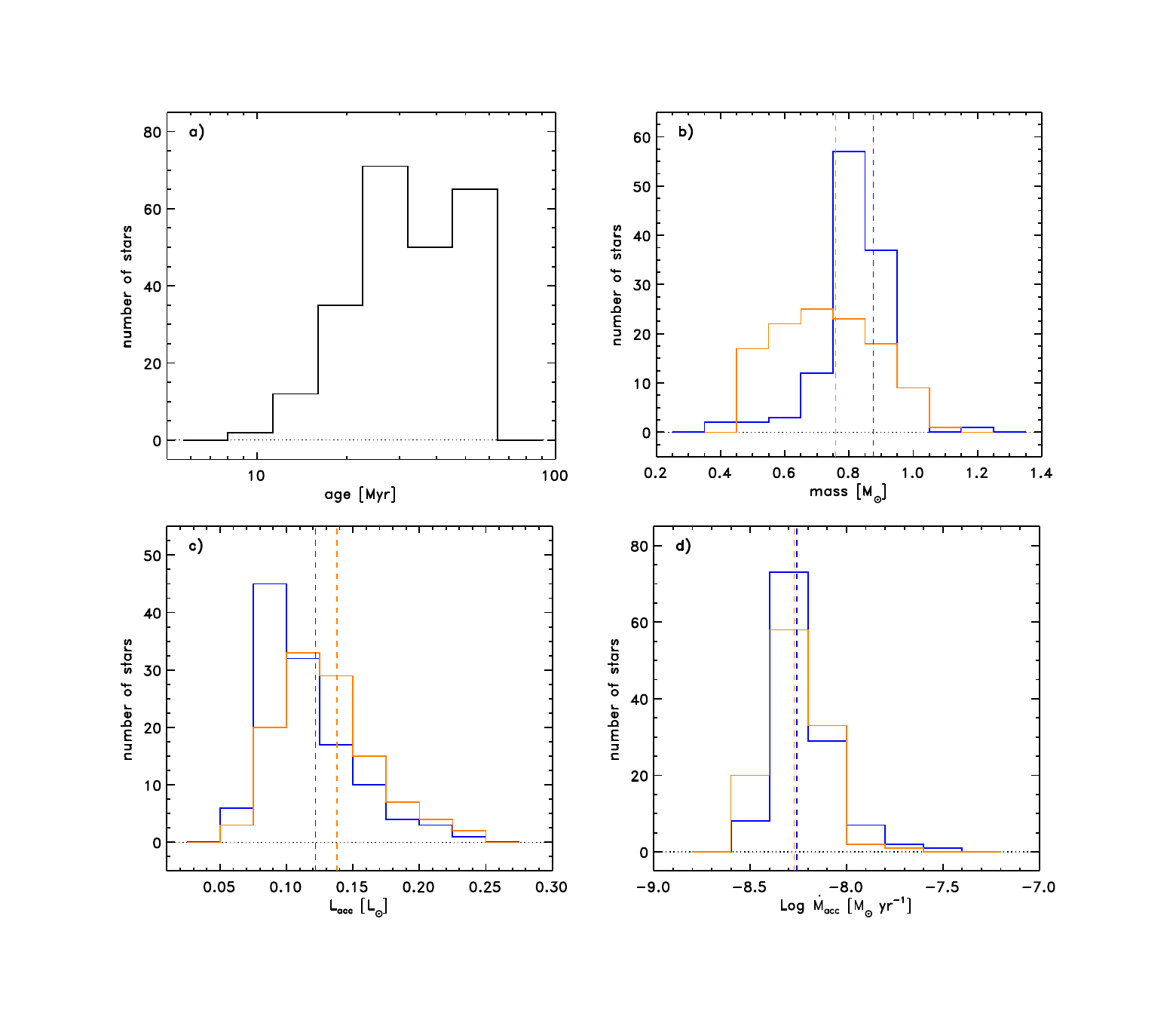}}
        \caption{Histograms showing the physical properties of the bona fide PMS stars, split into two age groups where relevant, indicated in blue for PMS stars younger than 32\,Myr and in orange for older objects. The panels show the distribution of the ages (panel a), masses (panel b), accretion luminosities (panel c), and mass accretion rates (panel d). The median value of the distributions for younger and older PMS stars are indicated by the dashed vertical line of the respective colour.}
        \label{fig:Histogram}
\end{figure*}

As mentioned before, the isochrones in Fig.\,\ref{fig:CMD} can give us an approximate idea as to the age of the PMS stars. To obtain a more accurate determination of the ages, we follow \citet{romaniello1998}, who proposed a statistical analysis of the positions of the stars in the CMD, or better Hertzsprung--Russell diagram (HRD), taking into account the uncertainties on the measured parameters. This method (hereafter the `Sieve') was later refined by \citet{demarchi2011, demarchi2011a, demarchi2013, demarchi2017}. It provides the probability distribution for each individual star to have a given value of the mass and age, given the specific photometric uncertainties affecting it. The method rests on the measurement errors alone and does not make any assumptions about the properties of the population, like for instance the functional form of the initial mass function.\\
\newline
Since the uncertainties on the measured parameters of the stars (effective temperature $T_{\rm eff}$ and bolometric luminosity $L$) depend on the accuracy of the photometry, we included in our sample only stars with a combined uncertainty in $V$ and $I$ of less than $0.1$ mag (see Sect. \ref{sec:photometry}). Furthermore, even though some differential extinction might be present, we showed that it is very limited in this field, as revealed by Fig. \ref{fig:Statsub}. Thus, ignoring it will not affect significantly the determination of masses and ages.\\
\newline
To follow the Sieve method, we first constructed the HRD. We derived the effective temperature $T_{\rm eff}$ from the de-reddened $V-I$ colours (only Milky Way foreground extinction was considered, as mentioned above), through interpolation in the \citet{bessell1998} models. We then computed the bolometric luminosity $L$ of the stars from the de-reddened magnitude in the $V$ band, adopting a distance modulus to the SMC of 18.91 $\pm$ 0.04 \citep{sabbi2011} and the model atmospheres from \citet{bessell1998}. The typical uncertainty on $T_{\rm eff}$ is of the order of 100\,K, while that on $L$ ranges between $5 - 10$\%.\\
\newline
The positions of the stars in the HRD are sampled using a uniform grid with evenly spaced cells and logarithmic steps in both temperature and luminosity. The sizes of the cells are set by the typical observational uncertainties. The cells are compared with the PMS evolutionary tracks of \citet{tognelli2011} for $Z = 0.004$ (as appropriate for the SMC), after interpolation to a fine mass grid with a logarithmic step of $0.025$. The most likely mass and age in each cell are computed from the time of permanence in the cell of all tracks that cross it. For any observed star, the cell of the grid to which that star belongs provides the most likely mass and age. In the comparison with these evolutionary tracks, binarity is not taken into account. The procedure is discussed in detail by \citet{demarchi2017}, to whom we refer the reader for further details.\\
\newline
The resulting ages and masses for the 240 PMS stars in our sample are shown by the histograms in Fig.\,\ref{fig:Histogram}\textcolor{blue}{a)} and \ref{fig:Histogram}\textcolor{blue}{b)}, respectively. The age distribution is binned in logarithmic steps of $\sqrt{2}$, which is a conservative estimate of the typical age uncertainty \citep{demarchi2011a}. In this diagram, peaks can be seen at the bin between 23 and 32 Myr as well as at the bin between 45 and 64, with a slight trough in between. This suggests that there might be two different age groups present in the cluster, with median ages around $\sim 25$\,Myr and $\sim $50\,Myr. This suggests that the region might have undergone two separate episodes of star formation in the recent past. Indeed, as already mentioned in Sect. \ref{sec:CMD}, we also see a similar split in the ages with the massive stars. We have also separated the two groups of younger and older PMS stars in the other panels of Fig. \ref{fig:Histogram}, with the split in ages placed at 32 Myr in consideration of our choice of age bins.\\
\newline
The mass distribution of the PMS stars is shown in Fig. \ref{fig:Histogram}\textcolor{blue}{b)}, where the histogram of PMS stars younger than 32 Myr is shown in blue, while that of older stars is indicated in orange. The median of each distribution is also indicated with a vertical dashed line of the respective colour. The identified PMS stars range in mass between $0.4$ and $1.2$\,M$_\odot$, with most stars having masses between $0.8$ and 1 M$_\odot$. From this diagram, it seems that the PMS stars in the younger group are generally more massive than the older ones, with the younger stars having an average mass of $0.87 \pm 0.11$ M$_\odot$ and the older stars having an average mass of $0.76 \pm 0.15$ M$_\odot$, although these values do overlap within their uncertainty. This is understood, from a physical perspective, because as time proceeds more of the massive PMS stars reach the MS and are no longer undergoing accretion and thus are no longer detectable as objects with $H\alpha$ excess emission. However, it is also possible that our mass distribution of the younger PMS stars is slightly skewed towards higher masses. In the CMD, potential young, low-mass PMS stars would be at colours redder than $V-I = 1$ (see also Fig. \ref{fig:CMD}), implying that they are more than $2.5$ times fainter in $V$ than in $I$, which would make them harder to detect. The steep drop in number of young PMS stars with masses smaller than $0.75$\,M$_\odot$ could be an indication of such a limitation in sensitivity in the $V$ band. \\

\begin{figure*}[ht]
        \centering
 
\makebox[\textwidth][c]{\includegraphics[scale=0.6]{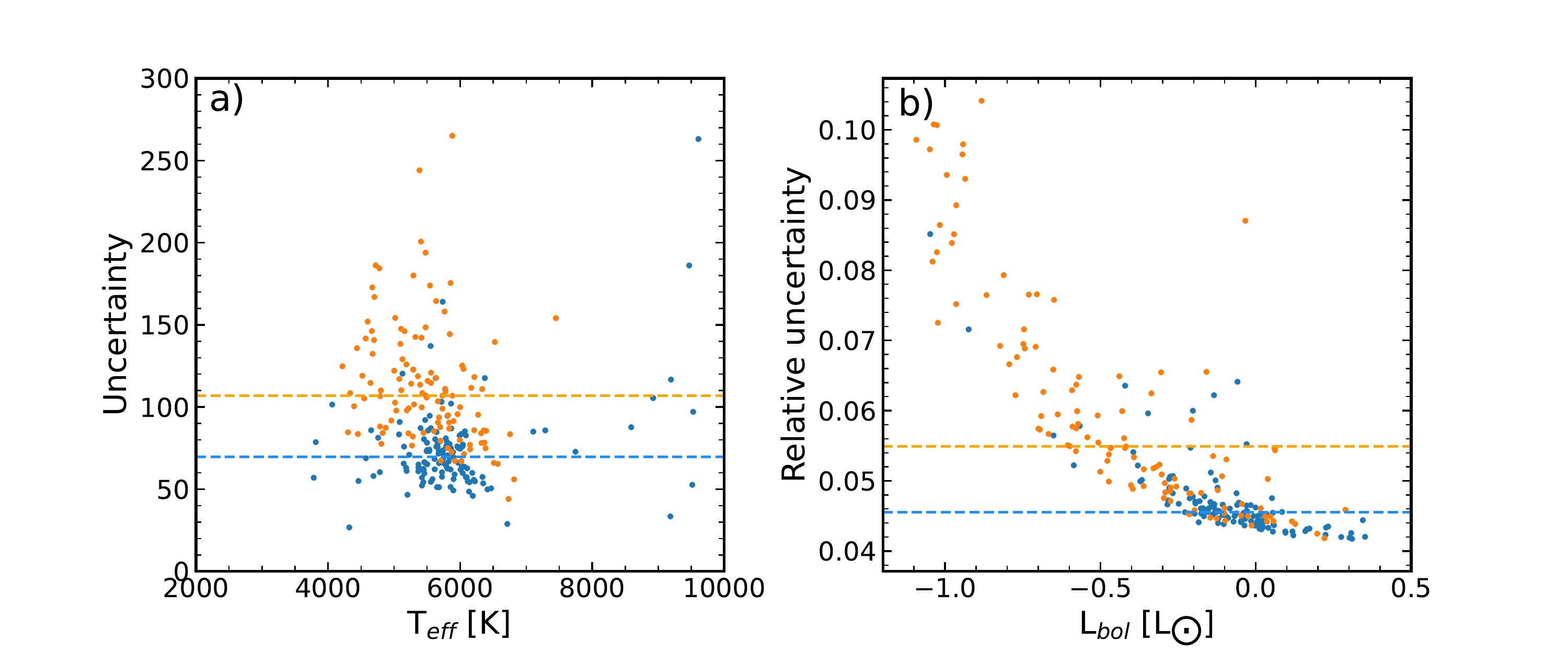}}
        \caption{Uncertainties on the $T_{\rm eff}$ (panel a) and $L_{\rm bol}$  (panel b). The PMS stars are divided into two age groups, with the younger objects indicated in blue and the older ones in orange. Dashed lines in these respective colours indicate the median (relative) uncertainty of each group.}
        \label{fig:LbolTeff}
\end{figure*}

\subsection{Accretion luminosity and mass accretion rate}\label{subsec:macc}

The accretion luminosity $L_{\rm acc}$ is derived from $L(H\alpha)$. Following \citet{demarchi2010}, $L(H\alpha)$ can be derived from the net excess emission $\Delta$H$\alpha$, already defined in Eq. \ref{eq:Delta}, the distance modulus to the SMC ($18.91 \pm 0.04$), as well as the sensitivity of the F656N filter (for further details, see \citealt{demarchi2010, demarchi2011, demarchi2013}). We can then relate $L(H\alpha)$ to $L_{\rm acc}$ with the following equation (see, e.g. \citealt{demarchi2010, demarchi2017}):\\
\begin{align}
        \label{eq:lacc}
        \log L_{\rm{acc}} = \left( 1.72 \pm 0.25 \right) + \log L(H\alpha).
\end{align}
\newline
Equation \ref{eq:lacc} makes the implicit assumption that there is a linear relationship between $L_{\rm acc}$ and $L(H\alpha)$. However, based on a purely empirical fit to the data, other authors have pointed out that the relationship is not necessarily linear and that the index can change with the value of $L(H\alpha)$ (see, e.g. \citealt{alcala2017}, and references therein). The uncertainty on the derived best fitting parameters in those works, however, is large. So, in the absence of a compelling reason for a non-linear relationship (see, e.g. \citealt{clarke2006}), we continued to adopt a linear relationship between the two quantities. Recently, \citet{tsilia2023} combined literature data (including those from the work of \citealt{alcala2017}) and derived a new linear relation between $L_{\rm acc}$ and $L(H\alpha)$ for a sample of 76 objects. The updated relationship that they obtain is $\log L_{\text{acc}} = \left( 1.38 \pm 0.25 \right) + \log L_{\text{H}\alpha}$, and therefore slightly less steep than the one of \citet{demarchi2010, demarchi2017}. However, within the uncertainties, the two slopes agree, so we continued to use the original relationship of \citet{demarchi2010, demarchi2017} to make it easier to compare our results with those of previous work on other star-forming regions studied in the same way.\\
\newline
Once the accretion luminosity, $L_{\text{acc}}$, is known, the mass accretion rate, $\dot{M}_{\text{acc}}$, can be calculated from the free-fall equation:\\
\begin{align}
        L_{\text{acc}} \simeq \frac{G M_*
\dot{M}_{\text{acc}}}{R_*}\left(1 - \frac{R_*}{R_{\text{in}}}  \right),
        \label{eq:freefall}
\end{align}
where $L_{\text{acc}}$ is the accretion luminosity of the star, $G$ is the gravitational constant, and $M_*$ is the stellar mass derived in Sect. \ref{subsec:mass_age}. $R_{\text{in}}$ is the inner radius of the circumstellar disk. Following \citet{gullbring1998}, we adopted $R_{\text{in}} = 5 R_*$. $R_*$ as the photospheric radius, which is derived from the photometry by comparing the absolute magnitude of the stars ($V$) with the theoretical values in the Bessell tracks ($V^{\text{ref}}$), where the radius is assumed to be 1\,R$_\odot$, using the adopted distance modulus $(m - M)_0 = 18.91 \pm 0.04$ as per the relationship
\begin{align}
    R_* = \sqrt{10^{-0.4\Delta V}}\,R_\odot, 
\end{align}
where $\Delta V = V - (m - M)_0 - V^{\text{ref}}$. \\
\newline
Histograms of the distribution of $L_{\text{acc}}$ and $\dot{M}_{\text{acc}}$ are shown, respectively, in Figs. \ref{fig:Histogram}\textcolor{blue}{c)} and \ref{fig:Histogram}\textcolor{blue}{d)}. Like before, blue and orange lines refer, respectively, to stars younger and older than 32 Myr. Even though, formally, stars younger than 32 Myr appear to have a slightly lower median $L_{\text{acc}}$, namely $0.12 \pm 0.04 L_\odot$ compared to $0.14 \pm 0.05 L_\odot$ for the stars older than 32 Myr, within the uncertainties no difference is present. No difference is present in the mass accretion rates either, and the average value of $\dot{M}_{\text{acc}}$ is $5.4 \times 10^{-9}$ M$_\odot$ yr$^{-1}$. All of these parameters are also summarised in Table \ref{tab:properties}. The positions, observed magnitudes, and derived physical parameters of all the 240 bona fide PMS stars are contained in Table \ref{tab:catalogue_PMS}, of which the first few lines are shown below. 

\subsection{Uncertainty on the parameters}\label{subsec:uncertainty}
Concerning the uncertainty on the derived physical parameters, we start from the effective temperatures and bolometric luminosities, from which ages and masses were also derived. We show in Fig. \ref{fig:LbolTeff}\textcolor{blue}{a)} and \ref{fig:LbolTeff}\textcolor{blue}{b)}, respectively, the absolute uncertainty on $T_{\rm eff}$ and relative uncertainty on $L_{\rm bol}$, as a function of the parameters. PMS objects younger and older than 32 Myr are indicated, as before, respectively in blue and orange, and the dashed lines correspond to the median values. The uncertainty on $L_{\rm bol}$ appears to be tightly correlated with $L_{\rm bol}$ itself and this can be understood because $L_{\rm bol}$ is directly derived from the $V$ magnitude, which shows a similar correlation with its uncertainty (see Fig. \ref{fig:Trumpet}). When considering the two age groups, we notice that the older population typically has a larger uncertainty on both parameters. This is not surprising, since we can see from Fig. \ref{fig:LbolTeff} that the older population is also typically fainter, in line with our earlier finding that older objects have typically lower mass (Fig. \ref{fig:Histogram}\textcolor{blue}{b}).  \\
\newline
As for the age and mass determination, this is limited both by the accuracy of the photometry and by the time resolution allowed by the isochrones, which decreases roughly exponentially as a PMS objects travels across the HRD (or CMD) along its evolutionary track. That means, for our method of age determination, that the closer a star approaches the MS, the less precisely can we know its age. Nonetheless, even though the ability to determine the absolute age of PMS stars is limited, we can still explore and characterise quite accurately any systematical differences in relative ages. In this sense, there might be evidence for the existence of two separate age populations within the cluster, as pointed out in Sect. \ref{subsec:mass_age}. For estimates on the typical uncertainties on the derived parameters, we refer to \citet{demarchi2017}. In that work, it was shown that the typical uncertainty of the age is about 18\,\%, and the typical uncertainty on the mass is approximately 6\,\%. It should be noted that the main source of uncertainty in \citet{demarchi2017} was the reddening. However, we have already shown in Sect. \ref{sec:CMD} that such an effect is likely negligible for NGC\,299, so these values can be considered as upper limits on the uncertainties on our parameters.\\
\newline
Finally, the uncertainty on the accretion luminosity and the mass accretion rate is dominated by the conversion from $L_{\text{H}\alpha}$ to $L_{\text{acc}}$ (Eq. \ref{eq:lacc}), which \citet{demarchi2011} have shown is no larger than $0.25$ dex, implying that the true intrinsic spread in the $L_{\text{H}\alpha}$ to $L_{\text{acc}}$ conversion cannot be larger than this. Thus, following \citet{demarchi2017} again, we estimate that the typical uncertainty on $L_{\text{acc}}$ and $M_{\text{acc}}$ is no more than 23\,\%. When comparing this finding to our results (Fig. \ref{fig:Histogram}\textcolor{blue}{d}), we can see that our data also show a spread of roughly $0.25$ dex. Thus, we are confident that this value provides a good measure of the general uncertainty of our measurements of the mass accretion rate.

\begin{table}[ht]
        \centering
        \begin{tabular}{c c c}
            \hline
             & Younger stars & Older stars  \\
             \hline
            Age [Myr] & 23 $\pm$ 7 & 47 $\pm$ 7 \\
            Mass [M$_\odot$] & 0.87 $\pm$ 0.11 & 0.76 $\pm$ 0.15 \\
            $L_{\text{acc}}$ [L$_\odot$]& 0.12 $\pm$ 0.04 & 0.14 $\pm$ 0.05 \\
            $\dot{M}_{\text{acc}}$ [M$_\odot$/yr] & $5.4 \cdot 10^{-9}$ &
$5.4 \cdot 10^{-9}$ \\
            \hline
        \end{tabular}
        \caption{All physical properties of the PMS stars derived in this
work, divided into their respective age groups. }
        \label{tab:properties}
\end{table}

\begin{sidewaystable}[h]
        \centering
        \begin{tabular}{c c c c c c c c c c c c c c c}
            \hline
            \hline
             ID & RA & Dec. & I &  V &  H$\alpha$ & T$_{\text{eff}}$ [K]
& L$_{\text{bol}}$ [L$_\odot$]& $R_*$ [R$_\odot$] & $M_*$ [M$_\odot$] &
$t$ [Myr] & $\delta t$ [Myr] & $L_{H\alpha}$ [L$_\odot$] & log
$\dot{M}_{\text{acc}}$ [M$_\odot$ yr$^{-1}$] \\
             \hline
            11 & 0:53:24.1842 & -72:13:51.023 & 24.50 & 25.80 & 22.35 & 4452 & 1.5E-01 & 0.84 & 0.61 & 44.8 & 6.7 & 2.8E-03 & -8.1\\
            30 & 0:53:26.4401 & -72:13:47.560 & 24.10 & 24.91 & 22.44 & 5559 & 3.4E-01 & 0.65 & 0.76 & 47.8 & 9.4 & 2.4E-03 & -8.4\\
            52 & 0:53:29.2780 & -72:13:44.602 & 25.14 & 26.31 & 22.22 & 4674 & 9.4E-02 & 0.56 & 0.57 & 54.6 & 9.4 & 3.4E-03 & -8.1\\
            137 & 0:53:29.2305 & -72:13:38.996 & 24.09 & 24.94 & 22.44 & 5418 & 3.3E-01 & 0.68 & 0.75 & 49.7 & 9.4 & 2.4E-03 & -8.3\\
            167 & 0:53:34.0782 & -72:13:37.190 & 24.70 & 25.67 & 22.34 & 5079 & 1.7E-01 & 0.58 & 0.66 & 57.3 & 9.4 & 2.9E-03 & -8.3\\
            214 & 0:53:33.6330 & -72:13:34.866 & 23.18 & 23.86 & 22.25 & 6018 & 9.0E-01 & 0.88 & 0.92 & 27.9 & 4.7 & 2.1E-03 & -8.4\\
            254 & 0:53:32.9969 & -72:13:32.793 & 24.80 & 25.61 & 22.39 & 5542 & 1.8E-01 & 0.47 & 0.68 & 53.8 & 9.4 & 2.8E-03 & -8.4\\
            264 & 0:53:39.4584 & -72:13:32.364 & 23.32 & 24.09 & 22.23 & 5694 & 7.3E-01 & 0.90 & 0.85 & 26.4 & 4.7 & 2.3E-03 & -8.3\\
            273 & 0:53:30.8870 & -72:13:31.992 & 23.02 & 23.59 & 22.17 & 6525 & 1.2E+00 & 0.84 & 1.00 & 36.3 & 6.7 & 2.2E-03 & -8.4\\
            335 & 0:53:37.0380 & -72:13:29.779 & 23.11 & 23.96 & 22.21 & 5421 & 8.2E-01 & 1.07 & 0.91 & 15.7 & 2.4 & 2.1E-03 & -8.3\\
            \hline
            \hline
        \end{tabular}
        \caption{Physical parameters of the 240 bona fide PMS stars.
Columns are: stellar ID, RA and Dec. position in the V band frame, magnitudes in V, I, and H$\alpha$ bands, effective temperature [K], bolometric luminosity [L$_\odot$], stellar radius [R$_\odot$], stellar mass [M$_\odot$], age [Myr], age uncertainty [Myr], H$\alpha$ luminosity [L$_\odot$], and mass accretion rate [M$_\odot$ yr$^{-1}$]. Only the first few lines are shown; the complete catalogue is available at the CDS.}
        \label{tab:catalogue_PMS}
\end{sidewaystable}

\subsection{Comparison with other SMC clusters}\label{subsec:comparison}
\begin{figure}[h]
     \centering
     \includegraphics[scale=0.60]{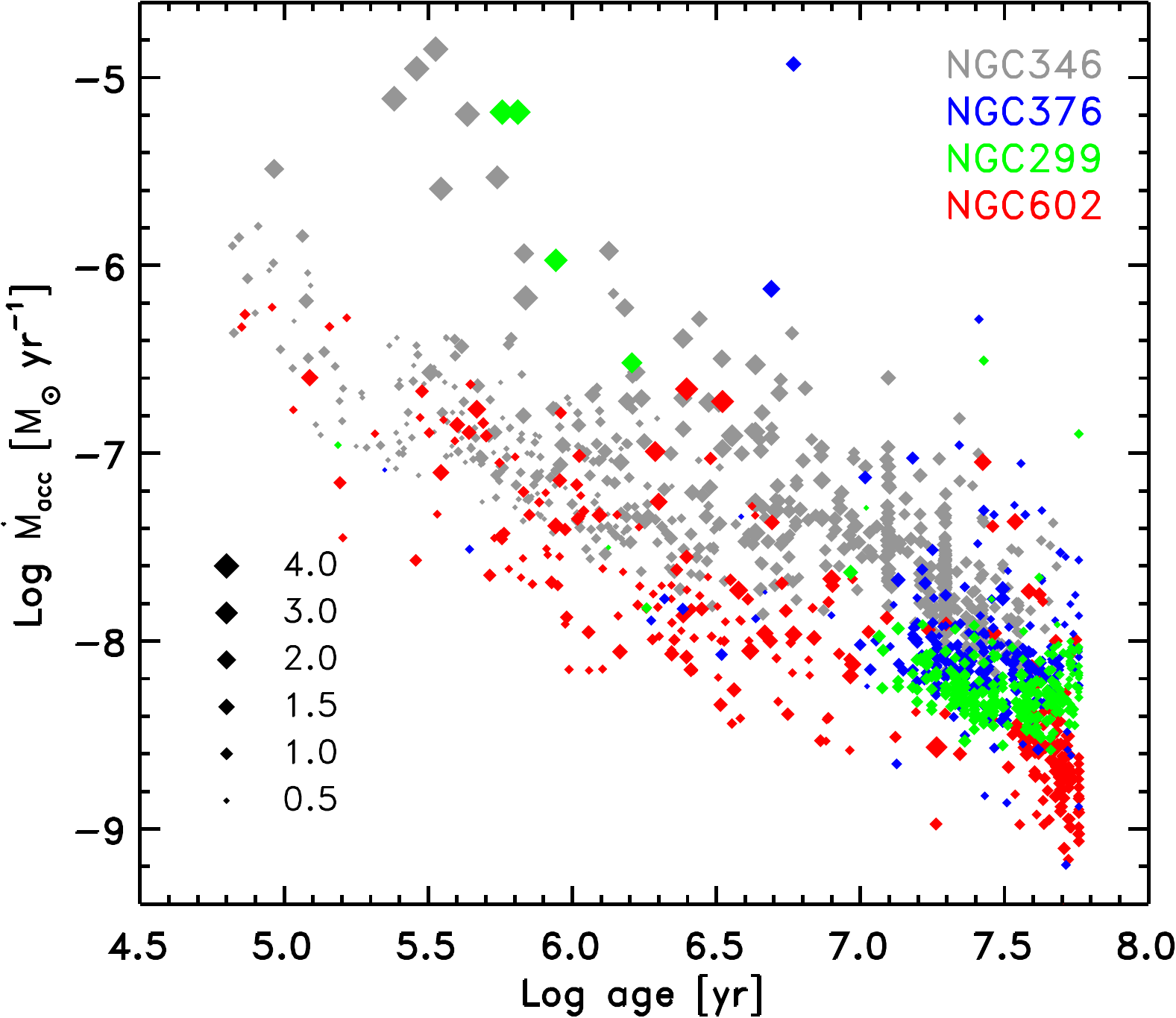}
     \caption{Ages and mass accretion rates for all PMS stars identified in NGC\,346 (\citealt{demarchi2011}, indicated in grey), NGC\,602 (\citealt{demarchi2013}, indicated in red), NGC\,376 (\citealt{tsilia2023}, indicated in blue), and NGC\,299 (this work, indicated in green).}
     \label{fig:macc_vs_age}
\end{figure}

The typical mass accretion rate of the PMS stars in NGC\,299, namely $5.4 \times 10^{-9}$ M$_\odot$ yr$^{-1}$ (or, $\log \dot{M}_{\text{acc}} = -8.3$), can be compared with the $\dot{M}_{\text{acc}}$ values measured in other clusters with metallicity similar to that of  NGC\,299. For example, \citet{demarchi2011} studied the cluster NGC\,346, located in the SMC. Considering objects in that work with ages and masses similar to those of our NGC\,299 sample, we find a typical value of $\log \dot{M}_{\text{acc}} = -7.9$ (see Fig. 10 in that paper). In a similar work, \citet{demarchi2013} studied NGC\,602 (also located in the SMC). From their Fig. 13, we derive a typical mass accretion rate $\log\dot{M}_{\text{acc}} = -8.2$, again considering only stars with masses and ages similar to those of our PMS stars in NGC\,299.\\
\newline
From the results on NGC\,346 and NGC\,602, combined with those stemming from a number of other regions that the same authors studied, \citet{demarchi2017} suggest that NGC\,602 has a lower characteristic mass accretion rate compared to NGC 346 and the other clusters, possibly because NGC\,602 is located in a lower-density environment. This possibility was recently explored by \citet{tsilia2023}, who studied the cluster NGC\,376 (also situated in a low-density SMC environment) and found it to have mostly PMS objects with an age of $\sim 30$\,Myr and a typical mass accretion rate $\log \dot{M}_{\text{acc}} = -8.2$. NGC\,299 is also located in a lower-density environment, so the typical mass accretion rate $\log \dot{M}_{\text{acc}} = -8.3$ that we measure appears to support the hypothesis suggested by \citet{demarchi2017} that the density of the environment, in addition to the age, mass, and metallicity of the PMS objects, can have an effect on the rate of mass accretion. Aside from calculating the average values for the mass accretion rate, we have also plotted the mass accretion rate as a function of the age for all of the PMS stars detected  in all four of these SMC clusters. This is shown in Fig. \ref{fig:macc_vs_age}. This clearly shows that the accretion rates found for the stars in NGC\,346 (indicated in grey) are typically higher than those found in NGC\,376 (blue), NGC\,602 (red) and NGC\,299 (green). Our hypothesis is further supported by the study of the PMS objects in the LMC clusters LH\,95 by \citet{biazzo2019} and LH\,91 by \citet{carini2022}. Both regions are characterised by a low-density environment, as witnessed by their mean dust density (see \citealt{carini2022}), and both reveal a systematically lower mass accretion rate for PMS stars in the range $0.4 - 1.0$\,M$_\odot$ than denser LMC regions such as 30 Dor. While still tentative, these results suggest that the gas density of the primordial molecular cloud could influence the typical size and mass of circumstellar disks that are formed in the cloud, which will then have an effect on the mass available for accretion onto the stars forming in that region.\\
\newline
Indications of the density of these environments are provided by existing surface density maps of the star-forming regions. \citet{utomo2019} use \textit{Herschel} observations to derive a map of the dust distribution and density of the SMC and LMC. These maps show that the dust surface density of the region containing NGC\,299 is approximately $0.003$\,M$_\odot$ pc$^{-2}$. We find a value of approximately $0.005$ M$_\odot$ pc$^{-2}$ for NGC\,376, a value of approximately $0.006$ M$_\odot$ pc$^{-2}$ for NGC\,602, and for NGC\,346 the value is $\sim 0.03$ M$_\odot$ pc$^{-2}$. This shows that the dust density in NGC\,346 is about an order of magnitude larger than that in NGC\,376, NGC\,602 and NGC\,299. One expects that, in general terms, the same ratio will apply to the gas density, as this scales directly with the dust density. The fact that the surface density of the dust in NGC\,299 is low is also supported by a visual inspection of the images, revealing a large number of background galaxies, suggesting that there is little dust attenuation in this area of the SMC. This is consistent with our previous analysis of the differential extinction in the field (Sect. \ref{sec:CMD}), which we found to be negligible, as also witnessed by the rather compact shape of the RC in the CMD (Fig. \ref{fig:CMD}). Thus, we can conclude, both in qualitative and quantitative terms, that the dust density and gas density in NGC\,299 is low. \\
\newline
It is reasonable to imagine that a low gas density in the area would imply that there is less gas present in the disks around forming stars. This could in turn cause the mass accretion rate to be low, but also possibly cause the accretion timescale to be short. However, when comparing multiple studies of PMS stars in SMC clusters, one finds that the accretion times appear to be similar for clusters in both high- and low-density regions, since similar ages are found for PMS stars in all of these studies. On the other hand, the mass accretion rate appears to be systematically lower in low-density regions \citep{demarchi2011, demarchi2013, tsilia2023}. Therefore, we conclude that the lower mass of the disks in low-density regions does not affect the duration of the accretion process, but rather its intensity.

\begin{figure*}[ht]
     \centering
    \makebox[\textwidth][c]{\includegraphics[scale=0.55]{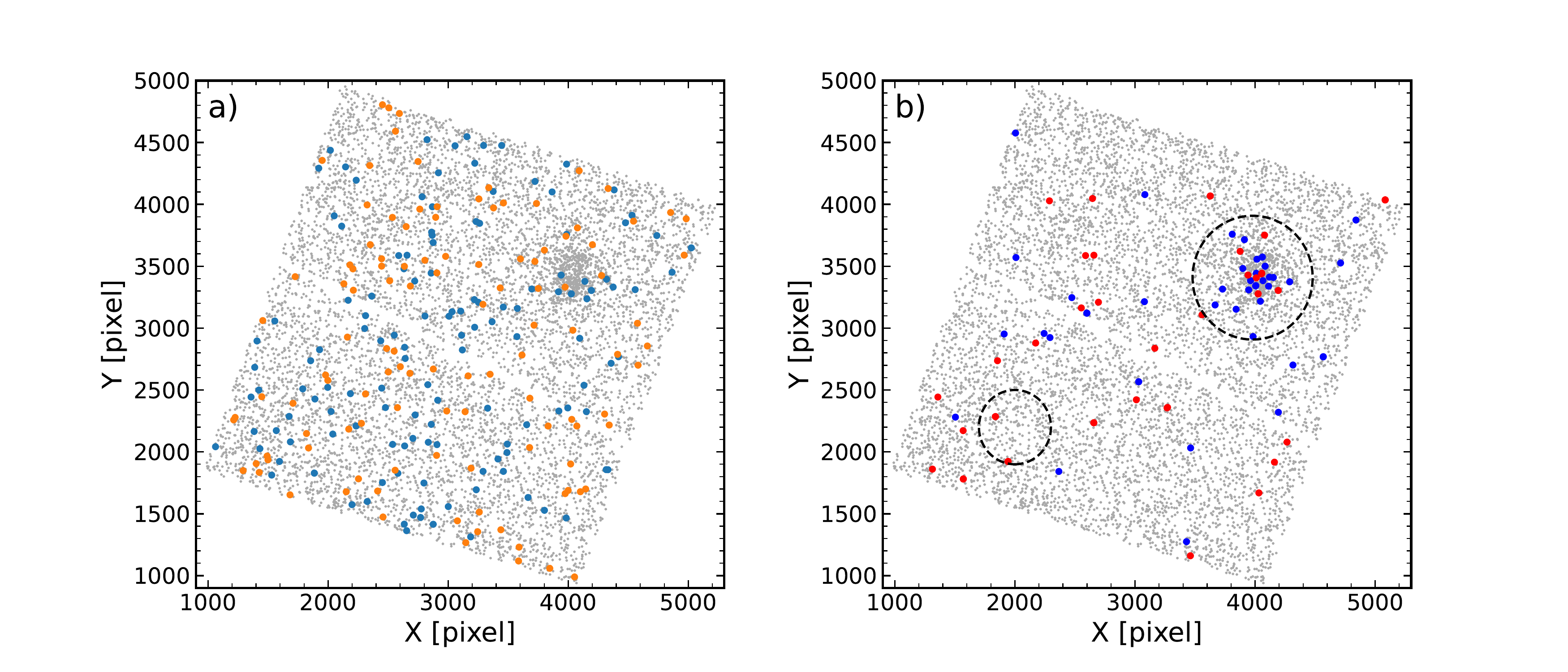}}
     \caption{Distribution of PMS stars (panel a) and massive stars (panel b) in the field. The younger PMS stars are indicated in blue and the older ones in orange. The younger massive stars are indicated in blue and the older ones in red. In panel b), two circles are drawn: one indicating a 25$\arcsec$ radius centred on the cluster, and one indicating the 15$\arcsec$ region used to estimate the background contribution to the stellar density profile. } 
     \label{fig:distr_PMS_massive}
\end{figure*}

\section{Dynamical state of NGC 299}\label{sec:dynamical}
\subsection{Spatial distribution of PMS stars and massive stars}\label{subsec:spatial}
\begin{figure}[h]
     \centering
     \includegraphics[scale=0.40]{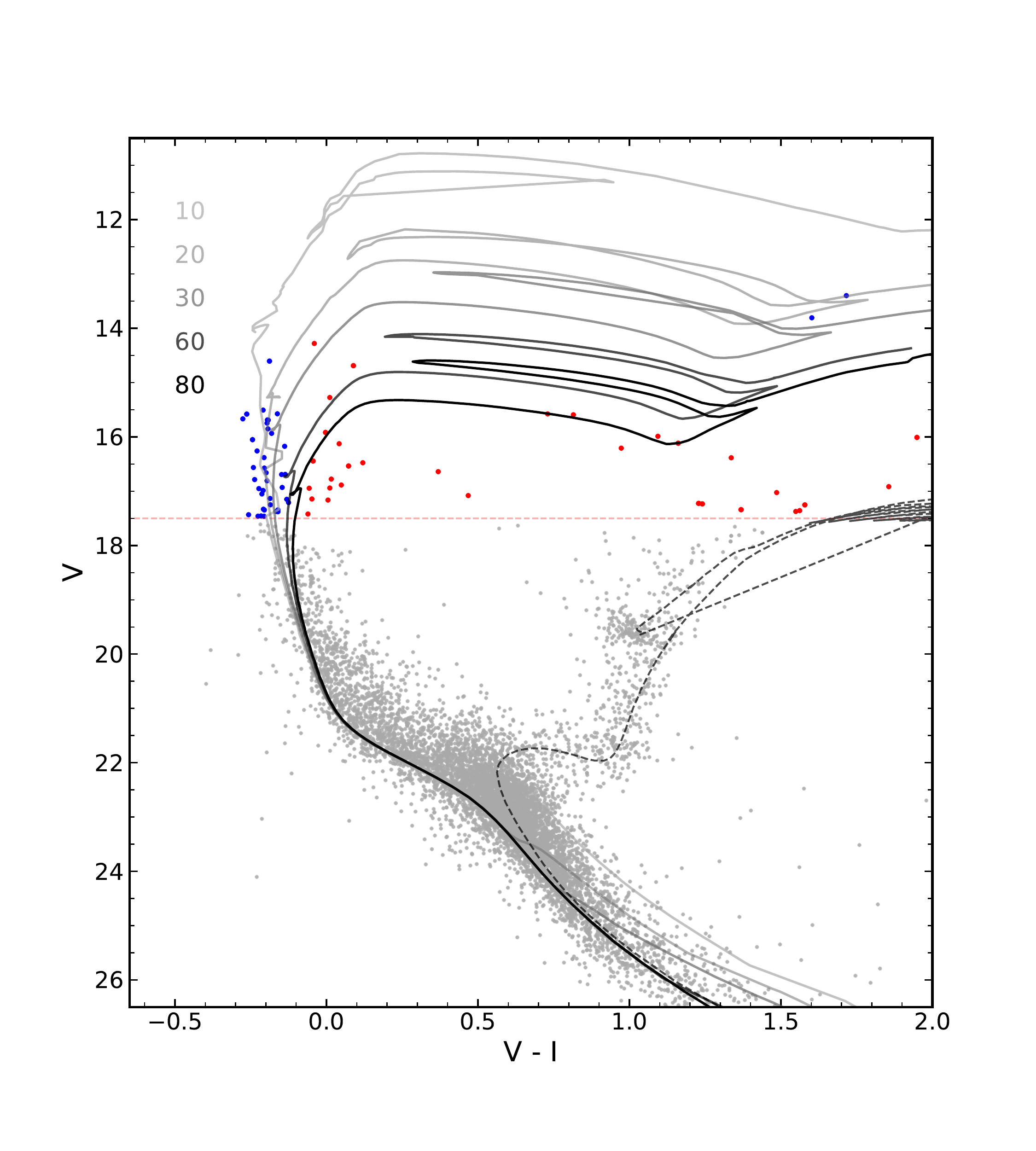}
     \caption{CMD showing the separation of the massive stars into two age groups. The complete sample of stars is indicated in grey, the younger massive stars are indicated in blue, and the older massive stars are indicated in red. Isochrones from the Padova group (see, e.g. \citealt{marigo2017}, and references therein) for ages of 10, 20, 30, 60, and 80 Myr are indicated in solid lines in shades of grey, with corresponding numbers in the image. Additionally, an isochrone for an age of 4 Gyr is indicated by a dashed black line.}
     \label{fig:CMD_massive}
\end{figure}

Our study of the PMS objects in NGC\,299 suggests that the cluster might have undergone two recent episodes of star formation, about 25 and 50\,Myr ago. The CMD in Fig. \ref{fig:CMD} reveals that massive stars also appear to be split in two age groups. Comparison with isochrones suggests an age of about 25\,Myr for the objects on the upper MS ($V\sim16$; $V-I \simeq -0.2$), while red supergiants ($V\sim16$; $0.7 \la V-I \la 1.3$) are more consistent with an age around 60--80 Myr. Despite the uncertainties inherent in the comparison with isochrones, the CMD shows that two recent star formation episodes are a likely possibility in the region of NGC\,299, with at least two populations of PMS and massive stars.\\
\newline
In Fig. \ref{fig:distr_PMS_massive} we show the spatial distribution of the two populations. Panel a) refers to PMS stars, where like in previous figures we indicate in blue the younger objects and in orange the older ones. As for the massive stars, shown in panel b), we selected objects brighter than $V=17.5$ (see also Fig. \ref{fig:CMD_massive}) and marked in blue the younger objects and in red the older ones. Although there is a small concentration of younger PMS stars near the centre of the cluster (upper right corner of the image), PMS stars are rather uniformly distributed, regardless of their age. Conversely, massive stars are clearly centrally concentrated, with the younger massive stars (indicated in blue) even more concentrated than the red supergiants. This difference can be clearly seen when considering the distributions of the younger and older massive stars both inside and outside of the 25$\arcsec$ radius that is drawn on the diagram. Outside the 25$\arcsec$ radius, both the younger and older massive stars are distributed roughly equally. However, within the 25$\arcsec$ radius, the younger massive stars are more than twice as numerous as the older massive stars. \\
\newline
All this suggests the presence of mass segregation in NGC\,299, since the low-mass PMS stars are much more widely distributed than massive objects. We do not know whether all the PMS stars that we detect formed inside the NGC\,299 cluster, but if they did, their current spatial distribution would not be unexpected. Indeed, young star clusters such as NGC\,299 can become mass segregated more quickly than one would expect from two-body relaxation processes alone if they formed from clumps of material that were already mass segregated by themselves (see, e.g. \citealt{mcmillan2007}). Furthermore, adopting a typical velocity dispersion of $\sim 1$ km\,s$^{-1}$, these stars can cross a distance of approximately one pc in 1 Myr. Our field of view covers about 50\,pc on a side, so with an approximate cluster age of $\sim 30$\,Myr one would expect the low mass stars (i.e. the PMS objects) to be fairly evenly spread throughout the field (see Fig. \ref{fig:distr_PMS_massive}). We note, however, that it is equally possible that these PMS star candidates were not more clustered in the past, but rather formed in a more diffuse manner throughout the cloud. In that case, even if these stars were clustered in very small structures, these would likely have dispersed very quickly so none of that would be visible today. \\
\newline
Concerning massive stars, the older ones have a much wider distribution than the younger massive objects. Since massive stars are likely to form in binaries and multiple systems \citep{mason1998, preibisch1999, garcia2001}, when one such star explodes as a supernova its companions will be ejected at high speeds. For the older population of massive stars, it is more likely that such a process has already occurred, so this could explain why they are more evenly distributed throughout the field and less centrally concentrated compared to their younger counterparts.

\subsection{Stellar density profile}\label{subsec:steldens}
\begin{figure}[ht]
     \centering
\includegraphics[scale=0.5]{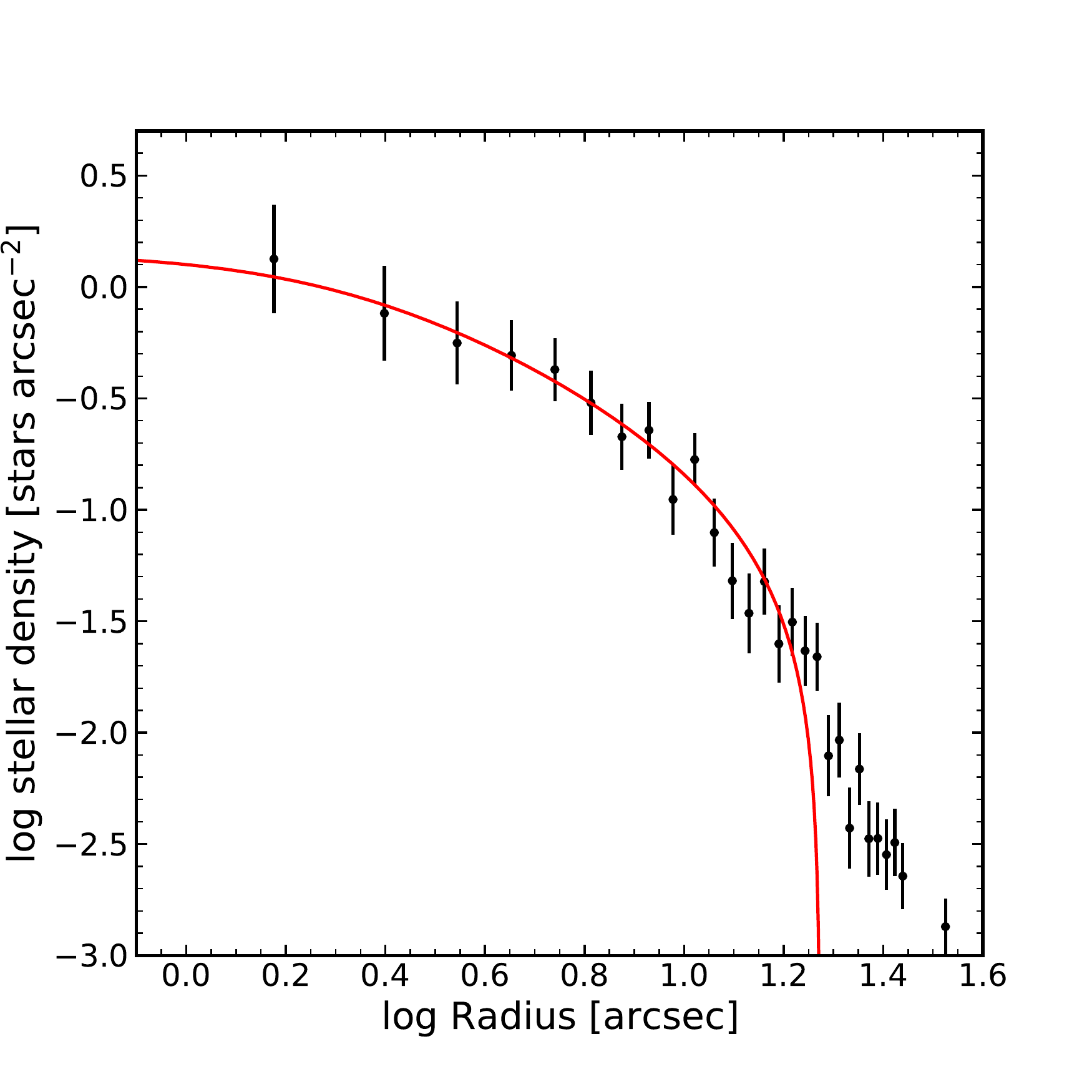}
     \caption{King model fit, indicated in red, to the stellar density profile of NGC 299, after subtraction of the field background. The error bars indicate the relative Poisson uncertainty on the measured number of stars in each annulus. }
     \label{fig:surf_stel_dens}
\end{figure}

If the wider distribution of PMS candidates is indeed the result of mass segregation, it could be a sign that NGC\,299 is currently dispersing into the field. To investigate the dynamical state of NGC\,299, we followed a similar procedure to what was done in \citet{sabbi2011}, who studied the SMC cluster NGC\,376 and found it to be dispersing into the field. We fitted a King model \citep{king1962} to the NGC\,299 stellar density profile in order to obtain a measure of the cluster's core radius, $r_c$, and tidal radius, $r_t$. The typical form of the King model is
\begin{align}
     \label{eq:kingrc}
     f = \frac{f_0}{1 + (r/r_c)^2,}
\end{align}
where $f_0$ is the central density. Equation \ref{eq:kingrc} can also be rewritten to include the tidal radius explicitly, which gives it the following form:
\begin{align}
     \label{eq:kingrcrt}
     f=k\left\{ \frac{1}{\left[1+(r/r_c)^{2}\right]^{1/2}} - \frac{1}{\left[1+(r_t/r_c)^{2}\right]^{1/2}}\right\}^{2},
\end{align}
where the $f_0$ and $k$ parameters are related as follows:
\begin{align}
     \label{eq:rtrc}
     f_0 = k \left\{1 - \frac{1}{\left[1 + (r_t/r_c)^2 \right]^{1/2} } \right\}^2.
\end{align}
\newline
The stellar density profile was obtained by dividing our image into equally spaced annuli centred on the cluster and summing over the number of stars in those annuli. To limit the effects of photometric incompleteness, only stars brighter than $V=24$ were used in this analysis. Contamination by field stars must be removed from our stellar density profile. This contamination was estimated to be approximately $0.016$ stars/arcsec$^2$, as measured in a small patch of sky $15\arcsec$ in diameter at the edge of our frame, at a projected distance of $115\arcsec$ from the cluster centre.\\
\newline
The best fit of the King model to the background-subtracted stellar density profile is shown in Fig. \ref{fig:surf_stel_dens}, corresponding to a core radius $r_c = 2.0\arcsec$, or $0.6$ pc, and a tidal radius $r_t = 18\farcs8$, or $5.5$ pc. The estimated uncertainty on $r_c$ is approximately $0\farcs5$ c and that on $r_t$ about $2\farcs5$.\\ 
\newline
As a reference, the core radius of NGC\,299 was determined once before in a study by \citet{hill2006}, who find a value of $r_c = $2.44$^{+1.23}_{-0.91}$ pc. This is larger than the value provided by our more accurate photometry (0.6 $\pm$ 0.15 pc), yet still compatible within the large uncertainty of \citet{hill2006}. The tidal radius of NGC\,299 had not been measured before.\\
\newline
The concentration parameter $c = \log_{10} (r_t/r_c)$ is routinely used to characterise the dynamical state of a cluster, that is, to establish whether a system is in virial equilibrium (e.g. \citealt{meylan1997}). In particular, systems with $c < 0.7$ are not in virial equilibrium and dispersion into the field is already advanced. Conversely, tightly bound systems, which describes most globular clusters, have values of $c \simeq 2$. With our constraints on the core radius ($2\arcsec$) and tidal radius ($18\farcs8$), we find that $c \simeq 1$ for NGC\,299, namely a concentration parameter around unity. Accordingly, the cluster would seem to still be in virial equilibrium, but the rather low concentration implies that the cluster might have already started to disperse into the field.\\
\newline
Finally, we can also tie this conclusion back to our findings for the two different age populations present in the cluster. We find that there was likely a previous star formation episode some 50 Myr ago, which, however, is hard to separate from the more recent episode on the basis of the masses and current accretion rates of the PMS stars. The finding that NGC\,299 could be dispersing into the field is compatible with the two populations being very similar by now, not only in regards to the accretion process, but also the spatial distribution.

\section{Summary and conclusions}\label{sec:conclusions}
In this work we study the cluster NGC\,299 in the SMC using observations obtained with the ACS and WFC3 instruments on board the HST. We searched for recently formed PMS stars in the field and determined their masses, ages, and mass accretion rates. We also compared our measure of the average mass accretion rate with that of several other SMC clusters to test the hypothesis that the density of the environment could influence the mass accretion process. Additionally, we studied the spatial distribution of the PMS stars and compared it to that of the massive stars in the cluster. Finally, we studied the dynamical state of the cluster by investigating its stellar density profile. The main results of this work can be summarised as follows.

\begin{enumerate}

\item
We constructed a $V$ versus $V-I$ CMD for the cluster and compared it to theoretical isochrones, from which we conclude that NGC\,299 is roughly 30 Myr old, confirming previous estimates of its age. From a statistical subtraction of the CMD of the field from that of the cluster, we conclude that extinction within the cluster, although present, is negligible for our purposes.

\item
From the $V-H\alpha$ versus $V-I$ CCD we detected 250 stars with $H\alpha$ excess emission above the 5\,$\sigma$ level and an EW in $H\alpha$ smaller than -20\,\AA, which we consider bona fide PMS stars. Ten of them are likely massive Be type stars, which exhibit $H\alpha$ excess due to stellar winds, and thus are not PMS stars.

\item
From a statistical analysis of the position of the remaining 240 bona fide PMS stars in the  HRD and comparison to theoretical isochrones, we derived the most likely ages and masses of these stars. In terms of age, there are possibly two populations of PMS stars, with median ages of $\sim 25$ Myr and $\sim$ 50 Myr.

\item
The average mass accretion rate of the PMS stars in NGC\,299 is $5.4 \times 10^{-9} $\,M$_\odot$ yr$^{-1}$. This value is comparable to that measured in NGC\,376 -- also an SMC cluster with similarly low metallicity and stellar density -- but systematically lower than the SMC cluster NGC\,346, which has similarly low metallicity but higher stellar density. This supports the hypothesis that lower-density regions might have systematically lower mass accretion rates for objects of otherwise similar mass, age, and metallicity.

\item
The spatial distribution of the PMS stars shows that they are spread rather uniformly throughout the field, with no discernible difference between the two age groups. Conversely, massive stars are more centrally clustered, exhibiting clear signs of mass segregation, with the younger population of massive stars being more concentrated than the older ones.

\item
We studied the cluster's dynamical state by fitting a King model to the stellar density profile. The best fitting parameters are  $r_c = 2\farcs0 \pm 0\farcs5$ (or $0.6$ pc) and $r_t = 18\farcs0 \pm 2\farcs5$ (or $5.5$ pc).

\item
We find the concentration parameter, $c$, to be around unity, suggesting that the cluster, while probably still in virial equilibrium, is likely already dispersing into the field.

\end{enumerate}
A better understanding of the accretion process in these low-metallicity environments is expected from the already conducted spectroscopic observations of PMS stars in NGC\,346 with the {\it James Webb} Space Telescope. A preliminary analysis of those observations \citep{demarchi2023} suggests longer timescales for the dissipation of circumstellar disks in low-metallicity environments than in the solar neighbourhood.

\begin{acknowledgements}
      We thank the anonymous referee for the helpful comments that helped improve this paper.
\end{acknowledgements}

%
%

\bibliographystyle{aa}
\bibliography{references}

\end{document}